\documentclass[12pt]{article}  
\usepackage{amsmath,amsthm,amsfonts,amscd,amssymb,eucal,latexsym}

\def\qed{\hfill $\square$ }
\def\ie{\textit{i.e.\ }}

\def\rest{\upharpoonright}

\def\cA{{\cal A}}
\def\cB{{\cal B}}
\def\cC{{\cal C}}
\def\cD{{\cal D}}

\def\cH{{\cal H}}

\def\cL{{\cal L}}

\def\cO{{\cal O}}

\def\cS{{\cal S}}
\def\cT{{\cal T}}
\def\cU{{\cal U}}
\def\cV{{\cal V}}

\def\NN{{\mathbb N}}

        \def\D{\Delta}
\def\eps{\varepsilon}

\def\r{\rho}
\def\s{\sigma}

\def\t{\tau}
\def\u{\upsilon}

\def\fA{{\mathfrak A}}

\newcommand{\bcA}{{\mbox{\boldmath$\cal A$}}}
\newcommand{\bcB}{{\mbox{\boldmath$\cal B$}}}

\newcommand{\bcS}{{\mbox{\boldmath$\cal S$}}}
\newcommand{\bcT}{{\mbox{\boldmath$\cal T$}}}
\newcommand{\bcU}{{\mbox{\boldmath$\cal U$}}}
\newcommand{\bcV}{{\mbox{\boldmath$\cal V$}}}

\newcommand{\br}{{\mbox{\boldmath $\rho$}}}
\newcommand{\bs}{{\mbox{\boldmath $\sigma$}}}

\newcommand{\bB}{{\mbox{\boldmath $B$}}}
\newcommand{\bR}{{\mbox{\boldmath $R$}}}
\newcommand{\bS}{{\mbox{\boldmath $S$}}}

\newcommand{\bU}{{\mbox{\boldmath $U$}}}
\newcommand{\bV}{{\mbox{\boldmath $V$}}}
\newcommand{\bW}{{\mbox{\boldmath $W$}}}
\newcommand{\ba}{{\mbox{\boldmath $a$}}}
\newcommand{\bx}{{\mbox{\boldmath $x$}}}
\newcommand{\bp}{{\mbox{\boldmath $p$}}}

\begin{document}

\title{Asymptotic Abelianness and Braided Tensor $C^*$--Categories}\smallskip

\author{Detlev Buchholz$^a$, Sergio Doplicher$^b$,  
Giovanni Morchio$^c$, \\ John E.\ Roberts$^d$ \hspace{0.3pt} and
\hspace{0.3pt} Franco Strocchi$^e$ \\[5mm]
\normalsize
$^a\,$Institut f\"ur Theoretische Physik der Universit\"at G\"ottingen,\\
\normalsize
D-37077 G\"ottingen, Germany\\ 
\normalsize
$^b\,$Dipartimento di Matematica, Universit\`a di Roma ``La Sapienza'',\\
\normalsize
I-00185 Roma, Italy \\ 
\normalsize
\hspace*{-9pt} $^c\,$Dipartimento di Fisica dell'Universit\`a and INFN,  
I-56126 Pisa, Italy,\\ 
\normalsize
$^d\,$Dipartimento di Matematica, Universit\`a di Roma ``Tor Vergata'',\\
\normalsize
I-00133 Roma, Italy\\ 
\normalsize
$^e\,$Scuola Normale Superiore and INFN, I-56126 Pisa, Italy}

\date{}
\maketitle

\begin{abstract} \noindent
By introducing the concepts of asymptopia and bi--asympto\-pia, 
we show how braided tensor $C^*$--categories arise in a natural way. 
This generalizes constructions in algebraic quantum field theory by 
replacing local commutativity by suitable forms of asymptotic Abelianness.
\end{abstract}

\vfill

\noindent 
{\small 
Supported in part by the MURST, CNR-GNAFA, INdAM-GNAMPA
and the European Union under contract HPRN-CT2002-00280}.\eject

\section{Introduction}

  The occurrence of superselection sectors in quantum theories 
was first recognized by Wick, Wightman and Wigner \cite{WWW} who 
gave two important examples of situations leading to sectors, 
the univalence rule and the electric charge. 
For a while it looked as if little would be touched by 
their fundamental discovery. The unrestricted superposition
principle had to be abandoned, but it at least remained valid within 
certain subspaces, the coherent subspaces. However the point
of view persisted that pure states of the theory were described by the
projective space associated to a preassigned Hilbert space, or 
rather a subset of that space to account for the new phenomenon.

But a drastic change in the basic picture allowed the new phenomenon to
spawn new ideas and results: as was stressed very early by 
Haag, in quantum field theory the local observables 
are fundamental and generate an algebra. The different sectors provide 
inequivalent irreducible realizations of that algebra  \cite{HK}. 
Furthermore, starting just from the vacuum sector and analyzing the 
structure of the algebra, we can in principle determine all sectors.

For superselection sectors describing localizable charges, this 
is possible by a selection criterion \cite{DHR2} singling 
out those irreducible representations.
The mathematical object that emerges and describes the structure of 
the sectors is a symmetric tensor $C^*$--category with 
conjugates and irreducible unit. 
It was later shown that any such category is isomorphic to a 
category of unitary representations of some compact group, unique 
up to isomorphism \cite{DR1}. Furthermore there is a canonical field net 
with ordinary Bose--Fermi commutation relations at spacelike
separations \cite{DR} where this group, the gauge group, acts as 
automorphisms with the original net as fixed--point net.

The above selection criterion does not 
select all the relevant representations in every case. Although weaker 
physically significant conditions covering a large class of 
theories with short range forces have been analyzed with success 
\cite{BF}, there is no known, or even plausible proposed criterion 
for singling out the relevant representations in all cases. 
In fact, it suffices to take the case of the electric charge, one of the key 
examples of \cite{WWW}, to realize that there are still unresolved problems. 
Essentially one would like in this case to arrive at a simple picture where 
the sectors are labelled by the electric charge corresponding to a gauge group 
$U(1)$. However for each value of the electric charge there are myriads of 
representations differing by their infrared clouds, cf.\ \cite{DS,BD} 
and references therein. To find such a simple picture one would 
either have to take equivalence classes or choose a suitable subset of
representations. Previous work 
on the problem of sectors in quantum electrodynamics include 
\cite{Bu,BDMRS1,BDMRS2,Fr,FMS1}.

  This paper has been a result of our attempts to describe the sector 
structure of quantum electrodynamics and of similar models exhibiting
long range effects. In view of the central role played 
by the symmetric tensor $C^*$--category, we here propose a 
method of constructing such categories which might prove to be applicable 
in these cases but the scheme seems of interest in its own right.

In the simplest case of strictly localizable charges \cite{DHR2}, 
one passes from the selected representations and their intertwiners 
to endomorphisms and their intertwiners using Haag duality. 
The endomorphisms of a $C^*$--algebra and their intertwiners form 
a tensor $C^*$--category, the endomorphisms being 
the objects and the intertwiners the arrows. The symmetry 
properties of the category can be deduced from analyzing the 
commutation properties of intertwiners. 

  The proposal described here allows for intertwiners 
not lying in the algebra where the endomorphisms act. In addition, our 
endomorphisms are not required to be locally inner but only asymptotically 
inner. We will show in Section 2
how an appropriate form of asymptotic Abelianness allows 
one to construct a tensor $C^*$--category from this structure
without the additional input of Haag duality.
Similarly, the symmetries on our derived tensor $C^*$--categories,
discussed in Section 3, 
will reflect asymptotic rather than purely local commutation 
properties of intertwiners; yet they are {\it exact symmetries} although 
one might have anticipated that asymptotic Abelianness would just 
lead to an asymptotic notion of symmetry.

The general mechanism presented here might still be too limited to be 
directly applicable to quantum electrodynamics, 
but might elucidate some important aspects of that theory.
As a matter of fact, it covers examples of theories exhibiting 
long range effects which go beyond the limits of the 
approaches previously studied. 

One such example is provided by the model 
of charges of electromagnetic type discussed in \cite{BDMRS1}, which 
actually stimulated the study of the more general structure discussed in 
the present paper. We briefly discuss at the end of Section 4 
how that model fits into the scheme presented here. %\smallskip

Another interesting example is concerned with the superselection 
structure of Bose sectors of the gauge invariant part of the algebra of a 
free  massive Fermi field on the two-dimensional space-time. Those sectors 
are usually described by a tensor category of localized automorphisms with 
a trivial Bose symmetry. The breakdown of Haag duality in this model, 
however, allows for different descriptions: for each real parameter 
$\lambda$, there is a tensor category describing the same sectors, where 
the intertwiners do not belong to the algebra, but fulfil the asymptotic 
Abelianness condition of Section 2. That category is equipped with a 
braiding which is not a symmetry, unless 
$\lambda = 0 $, and arises from a bi-asymptopia 
as described in Section 3. For $\lambda = 0 $,  one gets back the usual 
symmetric tensor category of localized automorphisms \cite{S}.

\section{Asymptotically Abelian Intertwiners}

\bigskip
One of the very first steps in the theory of superselection sectors is to 
show how a tensor $C^*$--category may be obtained by passing from a 
$C^*$--category of representations to a $C^*$--category of endomorphisms. 
In this step duality plays a fundamental role.
The aim here is to describe an alternative mechanism, 
in suitable  mathematical generality and abstraction.

   Let $\cA\subset \cB$ be an inclusion of unital $C^*$--algebras and 
$\Delta$ a semigroup of endomorphisms of $\cA$. Given $\rho,\sigma\in \Delta$,
we consider the corresponding space $(\rho,\sigma)$ of intertwiners 
$$(\rho,\sigma)=\{R\in \cB : R\rho(A)=\sigma(A)R, A\in \cA\}$$
and obtain in this way a $C^*$--category $\cT$, where the 
composition of arrows (intertwiners) is denoted by $\circ$.

This framework is supposed to model the situation of a set of 
representations and their intertwiners. 
Restricting attention to representations described by endomorphisms
does not seem restrictive on the mathematical side. 
In the case of separable simple $C^*$--algebras 
$\cA$ acting irreducibly on some Hilbert space $\cH$, 
any other irreducible representation can be obtained (up to 
equivalence) by the action of some automorphism.
One can choose that automorphism to be asymptotically inner in the
sense that it is the limit of inner automorphisms induced by a 
continuous one parameter
family of unitaries of the same algebra (cf. \cite{KOS}, and references in
there for previous results). 
If the algebra is also nuclear and not type $I$, any
cyclic representation is obtained in a similar way by asymptotically
inner endomorphisms \cite{K}. In the non separable case, customary 
in quantum field theory, the physically motivated split property \cite{DL}
implies that $\cA$ 
is generated by a type $I_\infty$ funnel. Then all
locally normal irreducible representations can also be 
described (up to equivalence) in one such representation by  
automorphisms \cite{Ta}. 

The intertwiners between the representations
considered, however, do not belong to the
given $C^*$--algebra in general, so End\,$\cA$ does
not model the category of representations, in
contrast to the case of localizable charges
in quantum field theory fulfilling duality. 
If duality fails and $\cB$ denotes the $C^*$-algebra
generated by the dual net, our category $\cT$
models again a category of representations with localizable charges. 
More generally, whenever the above mentioned theorems apply and $\cA$ acts
irreducibly on $\cH$, we may always take $\cB =  \cB(\cH)$. We take these
facts as motivation for studying our more general structure. 

The descriptions of representations given by
the above mentioned general mathematical
results, however, either ignore or take only partially into
account the local structure of $\cA$ in quantum field theory. Yet 
certain aspects of locality are needed in order to turn $\cT$ into
a \textit{tensor} $C^*$--category. Note that $\cT$ does not
yet have a tensor structure, for the arrows lie in $\cB$ and each 
$\rho \in \Delta$ is defined only 
on $\cA$.\footnote{$\cT$ acquires a tensor structure if
the tensor product of endomorphisms is defined by composition and 
if the tensor product of intertwiners $R\in(\rho,\rho')$ and
$S\in(\sigma,\sigma')$ given by $R\times S \doteq
R\rho(S)$ is meaningful as an element of $(\rho\sigma,\rho'\sigma')$.}
So what is required
is an extension of $\Delta$  to a semigroup of endomorphisms of 
the $C^*$--algebra  $\cA_\Delta\subset\cB$ generated by $\cA$ and all the 
intertwiner spaces $(\rho,\sigma)$ in $\cB$.

Two basic ingredients will help us to accomplish this task.
First, we assume that any $\rho\in\Delta$ is asymptotically inner 
in $\cB$, i.e.\ there are unitaries $U_m\in\cB$ such that 
\begin{equation} \label{1}
U_m^*AU_m^{}\to\rho(A) \end{equation}
in norm as $m\to\infty$ for each $A\in\cA$. 
Secondly, letting $V_n$ denote the analogues of the $U_m$ for 
$\sigma$ instead of $\rho$, equation (\ref{1}) implies that 
for all $R\in(\rho,\sigma)$, $A\in\cA$ 
$$[V_n^{}RU_m^*,A]\to 0$$ 
in norm as $m,n\to\infty$. 
As we want to extend our morphisms to $\cA_\Delta$, 
we require that this asymptotic Abelianness holds for intertwiners too, i.e.\ 
given $R\in (\rho,\sigma)$ and $R'\in (\rho',\sigma')$ 
\begin{equation}  \label{2}
[V_n^{}RU_m^*,R']\to 0 \end{equation}
in norm as $m,n \to \infty$. 
Indeed, condition (\ref{2})
 will allow us to use (\ref{1}) as a definition of the 
desired extension of each $\rho\in\Delta$ to $\cA_\Delta$.

  Rather than spell out our assumptions in detail, we will first introduce 
some standard formalism from the theory of $C^*$--algebras enabling us to 
replace a sequence of unitaries in $\cB$ by a unitary in a 
larger $C^*$--algebra 
$\bcB$. This simplifies what are already simple proofs by eliminating 
the indices. %\smallskip 

  Now the bounded sequences in $\cB$ with pointwise 
operations and norm defined 
as the supremum of the norms of elements of the sequence form a $C^*$--algebra 
and the subset of sequences that tend to zero in norm is a two--sided ideal 
in that algebra. $\bcB$ is then the quotient $C^*$--algebra and, 
passing from an element of $\cB$ to the corresponding constant sequences, 
$\cB$ can and will be canonically identified with a $C^*$--subalgebra
of $\bcB$, denoted again by $\cB$. 

We will use the following notation:
generic elements of $\bcB$ will be denoted by bold face letters $\bB$, 
whereas elements of $\cA, \cB$ and of their canonical 
images in $\bcB$ are denoted by $A,B$. 
As indicated, the elements  
$\bB \in \bcB$ are equivalence classes of bounded 
sequences modulo sequences that tend to zero. If  
$\{B_n\}_{n \in \NN}$ 
represents $\bB$ then $\|\bB\|=\limsup_n\|B_n\|$. Obviously any subsequence 
of a bounded sequence is again bounded and we say that a subset 
$\bcS\subset\bcB$ is {\it stable} if it is closed under taking 
subsequences.\smallskip 

\noindent
{\bf Lemma 1} {\sl A subset of $\bcB$ consisting of a single element 
$\bB$ is stable if and only if $\bB = B \in\cB$.}\smallskip 

\noindent
{\bf Proof.} Let $\{B_n \in \cB \}_{n \in \NN}$ be a sequence
representing $B \in \cB$, then  $B_n - B \to 0$. 
Given any subsequence $\{B_{n(i)}\}_{i \in \NN}$, 
then $B_{n(i)} - B \to 0$ and $\{B_{n(i)}\}_{i \in \NN}$
again represents $B$, so the subset consisting just of $\bB$ is stable. 
Conversely, if $\{B_n\}_{n \in \NN}$  does not represent an 
element of $\cB$, then $\{B_n\}_{n \in \NN}$ is not a Cauchy sequence. 
But then there is an $\eps>0$ and for each 
$i\in\NN$ an  $n(i),n'(i)\geq i$ 
with $\|B_{n(i)}-B_{n'(i)}\|\geq\eps$. 
Thus the subset consisting just of 
$\bB$ is not stable.\qed \smallskip

\noindent
{\bf Lemma 2} {\sl If $\bU$ is a unitary from $\bcB$ then there is a 
representing sequence consisting of unitaries.} \smallskip

\noindent
{\bf Proof.}  If $\{B_{n(i)}\}_{i \in \NN}$ is a representing 
sequence for $\bU$, then $B_n^*B_n^{}$ 
represents $\bU^*\bU=I$ so that 
$B_n^*B_n^{}\to 1$. Similarly $B_n^{} B_n^*\to 1$. 
Thus $|B_n|$ is invertible for all sufficiently large $n$ and if we then 
set $U_n \doteq B_n|B_n|^{-1}$, 
$U_n^*U_n^{}=1$ whilst 
$U_n^{}U_n^*=B_n^{}|B_n|^{-2}B_n^*$ is a projection tending 
in norm to $1$. Consequently, $U_n$ is unitary for all sufficiently 
large $n$ and, setting $U_n=1$ for other values of $n$ to make it 
unitary, the result follows. \qed
 
  Now if $\cU(\bcB)$ denotes the set of unitaries in $\bcB$, 
and $\bU\in\cU(\bcB)$, then 
$$\t (A) \doteq \bU^*A\bU,\quad A\in\cA$$ 
is a morphism of $\cA$ into $\bcB$ and we set 
$$\cU(\bcB)_\t \doteq \{\bU\in\cU(\bcB):\text{Ad}\bU^*\rest\cA=\t\}.$$ 

\noindent
{\bf Lemma 3} {$\t(\cA)\subset\cB$ if and only if $\cU(\bcB)_\t$ is 
stable.} \smallskip

\noindent
{\bf Proof.} If $\bU\in\cU(\bcB)_\t$ and 
$\cU(\bcB)_\t$ is stable then 
$\bU^*A\bU$ is stable and hence, by Lemma 1, 
 an element of $\cB$ for each $A\in\cA$. 
Conversely if $\t(\cA)\subset\cB$ then $\cU(\bcB)_\t$ is stable from the
definition. \qed \smallskip
 
Note that 
if we consider the $C^*$--category $\cC$ whose objects are the morphisms 
of $\cA$ into $\bcB$ inner in $\bcB$ and arrows their intertwiners 
then all objects are obviously unitarily 
equivalent since $\bU\in\cU(\bcB)_\t$ 
is a unitary in $\cC$ from $\t$ to the identity automorphism $\iota$. Now 
$\cA'\cap\bcB$ is the set of arrows from $\iota$ to $\iota$ in $\cC$ so 
if $\bU\in\bcU_\t$, $\bV\in\bcU_\u$ then 
$\bR\in(\t,\u)$ if and only if 
$\bV \bR \bU^*\in\cA'\cap\bcB$.

 Obviously if $\rho(\cA)\subset\cA$ so that we are dealing with endomorphisms 
of $\cA$ then $\cU(\bcB)_\r\cU(\bcB)_\s\subset\cU(\bcB)_{\s\r}$ 
for two such endomorphisms $\rho$ and $\sigma$. Now by Lemma 2, saying that 
an endomorphism of $\cA$ is unitarily implemented in $\bcB$ is the same 
as saying that it is asymptotically inner in $\cB$. So the set of such 
endomorphisms of $\cA$ obviously forms a semigroup. We begin here by studying 
a subsemigroup $\Delta$ and the associated $C^*$--category 
$\cT$ of intertwiners 
in $\cB$. Just as we defined $\bcB$ from $\cB$ so we can define a 
$C^*$--category $\bcT$ by  taking bounded sequences of arrows from 
$\cT$ with pointwise composition $\circ$ and 
the supremum norm and quotient by the ideal of sequences of arrows that tend 
to zero in norm. The proof of Lemma 2 shows that a 
unitary arrow of $\bcT$ 
has a representing sequence consisting of unitaries from 
$\cT$. $\cT$ can again 
be considered as a $C^*$--subcategory of $\bcT$ in a natural way.

   We now want to extend our semigroup of endomorphisms to the 
$C^*$--algebra $\cA_\D \subset \cB$ 
generated by $\cA$ and the intertwiners 
$\cT \subset \cB$ so that the extended endomorphisms remain 
asymptotically inner in $\cB$. The above discussion indicates what 
hypotheses will be necessary. We assume that for each 
$\r\in \D$, we are given a stable subset 
$\bcU_\r \subset \cU(\bcB)_\r$ with $\bcU_\r\cap\bcT\neq\emptyset$, where 
$\bcT$ here refers to the union of its Hom-sets. Thus in each 
$\bcU_\rho$ there is at least one sequence of intertwiners between morphisms 
belonging to $\Delta$ and 
$$\bcU_\r\bcU_\s\cap\bcU_{\s\r}\neq\emptyset.$$ 
Furthermore, we suppose that asymptotic 
Abelianness holds in the sense of Equation (\ref{2}), \ie 
given $R\in (\r,\s) \subset \cB$ and 
$R'\in (\r',\s') \subset \cB$ and $\bU\in\bcU_\r$, 
$\bV\in\bcU_\s$,
$$[\bV R\bU^*,R']=0.$$
Such a coherent 
assignment $\cU:\r\mapsto\bcU_\r$ for $\r\in\D$ will be 
called an {\it asymptopia} for $\D$.  Note that 
there are in general several such asymptopias. \smallskip

\noindent
{\bf Theorem 4} {\sl Let $\cA_\D$ denote the $C^*$--subalgebra of $\cB$ 
generated by $\cA$ and $\cT$, and let $\cU$ be some asymptopia for $\D$.
Then every $\rho\in \Delta$ has a unique extension 
$\rho_\cU$ to an endomorphism of $\cA_\D$ such that
$$\rho_\cU(A)=\bU^*A\bU,\,\,\,A\in\cA_\D\quad \bU\in \bcU_\rho.$$
Furthermore $(\r\s)_\cU = \r_\cU \s_\cU$ and 
$(\r,\s)=(\r_\cU,\s_\cU)$.
Thus $\cT$ inherits the structure of a tensor $C^*$--category from 
End\,${\cA}_\D$.}\smallskip

We first prove the following lemma.\smallskip

\noindent
{\bf Lemma 5} {\sl If $\rho,\sigma,\tau\in \Delta$ and 
$S\in (\sigma,\tau)$ then 
$\rho_\cU(S) \doteq \bU^*S\bU$ is independent of the choice of 
$\bU\in\bcU_\rho$
and is an element of $\cA_\D$.}\smallskip

\noindent
{\bf Proof.} If $\bU,\bU' \in \bcU_\rho$,
\begin{equation*}
\begin{split} 
& \bU^*S\bU - \bU'^* S\bU'=\bU^*(S\bU \bU'^{*} - \bU \bU'^{*}S)\bU' \\
& =\bU^* [S, \,\bU 1_\rho \bU'^{*}] \bU'=0 
\end{split}
\end{equation*}
as required. In particular, the subset consisting 
of the single element $\bU^*S\bU$ 
is stable since $\bcU_\rho$ is stable. 
Thus by Lemma 1, $\r_\cU(S)\in\cB$ and 
since we can choose $\bU\in\bcT$, 
$\r_\cU(S)\in\cA_\D$. \qed \smallskip 

\noindent
{\bf Proof of Theorem 4.} Lemma 5 shows that each $\r\in\D$ has 
the required 
unique extension to an endomorphism $\r_\cU$ of $\cA_\D$. The condition of 
asymptotic Abelianness shows that we do not lose any intertwiners in this way, 
\ie  that $(\r,\s)=(\r_\cU,\s_\cU)$. 
Since $\bV\bU\in\bcU_{\r\s}$ for some 
$\bU\in\bcU_\r$ and $\bV\in\bcU_\s$, 
$(\rho\sigma)_\cU = \r_\cU \s_\cU$. Thus
$\cT$ can be identified with a full tensor $C^*$--subcategory of 
End\,$\cA_\D$ 
completing the proof of the theorem. \qed \smallskip

   For a given semigroup $\Delta$ of endomorphisms, the 
$C^*$--category $\cT$ is determined 
by the inclusion $\cA\subset\cB$. Its tensor structure is determined 
by the choice of asymptopia 
$\cU$. Different asymptopias can lead to different tensor structures, in 
other words to different extensions of $\D$ to $\cA_\D$, unless 
they are included in a common asymptopia. 
$\r\mapsto\bcU_{\r_\cU}$, the set of all unitaries in $\bcB$ 
inducing $\r_\cU$  on $\cA_\D$, is obviously a maximal asymptopia 
containing the given 
asymptopia. Thus two asymptopias lead to the same tensor structure if
and only if there is 
an asymptopia containing both. 
If we consider the set of asymptopias ordered 
under inclusion, the different tensor structures correspond to the different 
path--components of this set.\smallskip

\noindent
{\bf Theorem 6} {\sl Given an inclusion of unital $C^*$--algebras $\cA\subset\cB$ and a 
semigroup $\Delta$ of endomorphisms of $\cA$, the path--components of the set of asymptopias 
for $\Delta$ are in natural 1--1 correspondence with the set of maximal asymptopias and with the 
set of extensions of $\Delta$ to  a semigroup of asymptotically inner endomorphisms of the 
$C^*$--algebra $\cA_\D$ 
generated by $\cA$ and the intertwiners for $\Delta$ in $\cB$.} \smallskip 

It is of interest that the above framework for studying
semigroups of endomorphisms can be extended so as to
cover the case of representations and their intertwiners.
One then deals with a set $\Delta$ of morphisms of $\cA$ into $\cB$ and the 
associated $C^*$--category of intertwiners. We suppose that for each 
$\rho\in\Delta$, we are given a set $\bcU_\rho$ of unitary operators 
in $\bcB$ with $\bcU_\rho\cap\bcT\neq\emptyset$ such that
$$\rho(A)=\bU^*A\bU$$
for each $A\in \cA$, $\bU\in \bcU_\rho$ and  
$\rho\in\Delta$. In particular, the inclusion mapping $\iota$ of $\cA$ into 
$\cB$ is in $\Delta$ with $\bcU_\iota$ consisting just of the identity
of $\bcB$. Furthermore, we require that for 
each $\rho,\sigma\in\Delta$, there is a unique $\tau\in\Delta$ such that 
given $\bU\in\bcU_\rho$ there is a 
$\bV\in \bcU_\sigma$ with $\bV \bU\in \bcU_\tau$. 
As before, $R\in (\rho,\sigma)$ if and only if $\bV R\bU^*\in\cB\cap\cA'$. 
The intertwiners  are again supposed to be asymptotically Abelian. 
Under these circum\-stances we have the following generalization 
of Theorem 4.
\smallskip 

\noindent
{\bf Theorem 7} {\sl Let $\cA_\Delta$ denote the $C^*$--subalgebra of $\cB$ 
generated by $\cA$ and $\cT$, then every 
$\rho\in \Delta$ has a unique extension 
$\rho_\cU$ to an endomorphism of $\cA_\Delta$ such that
$$\rho_\cU(A)=\bU^*A\bU,\,\,\,A\in\cA_\Delta\quad \bU\in \bcU_\rho.$$
Furthermore $(\rho,\sigma)=(\rho_\cU,\sigma_\cU)$ 
and the set $\D_\cU$ of the extended morphisms constitutes a  
unital semigroup of endomorphisms of $\cA_\Delta$.
Thus $\cT$ inherits the structure of a tensor $C^*$--category from 
End\,$\cA_\Delta$.}\medskip 

\noindent
{\bf Proof.} Note that Lemma 5 retains its validity in this new 
context and that $\rho(\cA)\subset\cA_\Delta$ since
$\bcU_\rho\cap\bcT\neq
\emptyset$. Thus $\rho_\cU(\cA) \doteq \bU^*\cA_\Delta \bU\subset\cA_\Delta$. 
Now let $\rho,\sigma\in\Delta$.  
Choose $\tau\in\Delta$, $\bU\in\bcU_\rho$ and $\bV\in\bcV_\sigma$ such that 
$\bV \bU\in\bcU_\tau$. 
Since Ad\,$\bU^*\bV^*$=Ad$\bU^*$Ad$\bV^*$, 
we conclude that $\rho_\cU\sigma_\cU$=$\tau_\cU$. Thus $\Delta_\cU$ 
is a semigroup with unit. Given $T\in(\rho,\sigma)$ then the set of $B\in\cB$ 
such that 
$$T\,\bU^*B\bU=\bV^*B\bV\,T,$$ 
$$T\,\bU^*B^*\bU=\bV^*B^*\bV\,T,$$ 
is a $C^*$--subalgebra of $\cB$ containing $\cA$. It also 
contains every arrow of $\cT$ since intertwiners are asymptotically Abelian.
This shows that $(\rho,\sigma)=(\rho_\cU,\sigma_\cU)$. \qed
\smallskip

\noindent
We conclude with the remark that all 
results of this section can be generalized  in the following sense. 
Instead of considering an inclusion of unital $C^*$--algebras, we can consider 
a fixed unital $C^*$--algebra $\cL$, say, and a set $\D$ of morphisms 
between unital $C^*$--subalgebras, closed under composition. Given 
$\rho,\sigma\in\Delta$, between two such subalgebras $\cA$ and $\cB$, we set 
$$(\rho,\sigma) \doteq \{R\in\cL:R\rho(A)=\sigma(A)R,\,\,A\in\cA\},$$ 
and obtain in this way a $C^*$--category $\cT$. 
In the presence of an asymptopia 
$\cU$ for $\Delta$, the morphism $\rho$ has a unique extension to 
a morphism $\rho_\cU$ between $\cA_\D$ and $\cB_\D$, the $C^*$-subalgebras 
of $\cL$ generated by $\cA$ and $\cT$ and $\cB$ and $\cT$, respectively. 
In this way, $\cT$ inherits the structure of a $2$--$C^*$--category from 
the $2$--$C^*$--category of morphisms and intertwiners between unital 
$C^*$--subalgebras of $\cL$, cf.\ \cite{LR}. The reader should have 
no difficulty in enunciating and proving the analogue of the results 
of this section.%\smallskip 

  The above generalization may prove relevant to the theory of 
superselection sectors first because it may prove advantageous to restrict 
a representation to a subnet of the observable net and secondly because 
it may not be possible to get an endomorphism of the subnet by passing 
to an equivalent representation. 

\section{The Emergence of Braiding}
 
In view of the results in the 
preceding  section, we can now forget about the
inclusion $\cA\subset\cB$, work with a single unital $C^*$--algebra $\cA$
(corresponding to the previous algebra $\cA_\D$) and suppose that
we are given a tensor $C^*$--category $\cT$ with $\Delta$ as its set
of objects realized as a full tensor subcategory of
End\,$\cA$. How it was obtained, in
particular, which asymptopia, if any, was used to construct it, is for the
moment quite irrelevant. 

We also note that the $C^*$--category $\bcT$
now becomes a tensor $C^*$--category in a natural way. The tensor
product of objects is just the pointwise product 
of the individual sequences    
of endomorphisms whilst if arrows $\bR$ and $\bS$ are represented by
sequences
$R_n\in(\rho_n,\rho'_n)$ and $S_n\in(\sigma_n,\sigma'_n)$, their
tensor product $\bR \times \bS$ is represented by
$R_n\rho_n(S_n)\in(\rho_n\sigma_n,\rho'_n\sigma'_n)$, $n \in \NN$. 
We further note that a sequence of endomorphisms 
$\r_n$ of $\cA$ defines in a canonical way
a unital morphism $\br$ of $\cA$ into $\bcA$, two sequences $\rho_n$
and $\sigma_n$ defining the same morphism if and only if
$\rho_n(A)-\sigma_n(A)\to 0$ for every $A\in\cA$. Adjoining
intertwiners, we get a $C^*$--category Mor\,$(\cA,\bcA)$. There is   
now an obvious canonical $^*$--functor $F$ from $\bcT$ to
Mor\,$(\cA,\bcA)$ mapping a sequence of endomorphisms of $\cA$ 
onto the induced morphism from $\cA$ to $\bcA$ and acting as 
the identity on intertwiners. The effect of 
$F$ is to identify sequences of endomorphisms
with the same asymptotic behaviour.

  The task in this section is to show how to get a braiding and
to develop criteria for deciding whether the braiding is a symmetry.
 The idea here is to define the braiding using suitable norm convergent
sequences of unitaries and we will again realize these sequences as 
unitary
operators in $\bcA$. However, the conditions we need are now somewhat
different. In particular, as a braiding is a function of two objects, we
assume that there are for each object $\r \in \D$ two stable sets of unitaries
$\bcU_\r , \bcV_\r$ 
which are contained in $\bcT$,  
each unitary $\bW$ implementing $\rho$, \ie
$\rho(A)=\bW^*A\bW$, $A\in\cA$. This means that $F(\bW)$ is an intertwiner in
$\bcA$ from $\rho$ to the identity automorphism $\iota$.
Furthermore, we require the following notion
of asymptotic Abelianness, where we work in the category 
Mor\,$(\cA,\bcA)$: given two intertwiners $R\in(\rho,\rho')$ and
$S\in(\sigma,\sigma')$ of $\cT$ and $\bU\in \bcU_\rho$, 
$\bU'\in\bcU_{\rho'}$,
$\bV\in\bcV_\sigma$ and $\bV'\in\bcV_{\sigma'}$ then
\begin{equation} \label{3}
F(\bU'R\bU^*\times \bV'S\bV^*)-F(\bV'S\bV^*\times \bU'R\bU^*)= 0.
\end{equation}

\noindent 
This form of asymptotic Abelianness of the tensor product of arrows will allow
us to define the universal rule for interchanging tensor products, in 
very much 
the same way as condition (\ref{2}) allowed us to define the tensor product 
of arrows.

For that purpose this data should, in addition, 
be compatible with 
products in the sense  that given $\rho,\rho'\in\Delta$, we can find
$\bU\in\bcU_\rho$ and $\bU'\in\bcU_{\rho'}$ such that
$\bU\times \bU'\in\bcU_{\rho\rho'}$ and analogously for 
the second set of unitaries $\{ \bcV_\r \}$. 
Any assignment $(\cU,\cV) : \r \mapsto \bcU_\r, \bcV_\r$
with these  properties will 
be referred to as a {\it bi-asymptopia} for $\D$. \smallskip

% \pagebreak
\noindent
{\bf Theorem 8} {\sl Let $(\cU,\cV)$ be a bi-asymptopia for $\D$. 
Given $\rho,\,\sigma\in\Delta$, then
$$\varepsilon(\rho,\sigma) \doteq F(\bV^*\times \bU^*)\circ F(\bU\times \bV)$$
is independent of the choice of $\bU\in \bcU_\rho$ and $\bV\in\bcV_\sigma$
and is in $(\rho\sigma,\sigma\rho)$.
Furthermore, if $R\in (\rho,\rho')$ and $S\in (\sigma,\sigma')$ then
$$\varepsilon(\rho',\sigma')\circ R\times S=S\times R\circ 
\varepsilon(\rho,\sigma)$$\smallskip
and if $\tau\in\Delta$ then
$$\varepsilon(\rho\sigma,\tau)=\varepsilon(\rho,\tau)\times 1_\sigma\circ 
1_\rho\times\varepsilon(\sigma,\tau),$$
$$\varepsilon(\rho,\sigma\tau)=1_\sigma\times \varepsilon(\rho,\tau)\circ 
\varepsilon(\rho,\sigma)\times 1_\tau.$$ 
In other words, $\varepsilon$ is a braiding for the full subcategory of 
End\,$\cA$ generated by $\Delta$.}\smallskip

\noindent
{\bf Proof.}
\begin{equation*} 
\begin{split}
& F(\bV^*\times \bU^*)\circ F(\bU\times \bV)-
F(\bV'^{*}\times \bU'^{*})\circ F(\bU'\times \bV') \\
& = F(\bV^*\times \bU^*)\circ \big(F((\bU\circ\bU'^{*})\times 
(\bV\circ \bV'^{*})) \\ 
& - F((\bV\circ \bV'^{*})\times (\bU\circ \bU'^{*}))\big)\circ F(\bU'\times \bV')=0
\end{split}
\end{equation*}
proving independence of the choice of
$\bU\in \bcU_\rho$ and $\bV\in\bcV_\sigma$. But then 
$F(\bV^*\times \bU^*)\circ F(\bU\times \bV)$
is stable so that $\varepsilon(\rho,\sigma)\in(\rho\sigma,\sigma\rho)$.
Furthermore,
\begin{equation*} 
\begin{split}
& F(\bV'^{*}\times \bU'^{*})\circ F(\bU'\times \bV')\circ R\times S \\
& - S\times R \circ F(\bV^*\times \bU^*)\circ F(\bU\times \bV) \\
& = F(\bV'^{*}\times \bU'^{*})\circ \big(F(\bU'R\bU^*\times \bV'S\bV^*) \\
& - F(\bV'S\bV^*\times \bU'R\bU^*)\big)\circ F(\bU\times \bV)=0,
\end{split}
\end{equation*}
and we deduce that
$$\varepsilon(\rho',\sigma')\circ R\times S=S\times 
R\circ\varepsilon(\rho,\sigma).$$%\smallskip
Now, pick $\bU\in \bcU_\rho$ and $\bV\in\bcU_\sigma$ such that
$\bU\times \bV\in\bcU_{\rho\sigma}$. Then, if $\bW\in\cV_\tau$,
$$F(\bW^*\times \bU^*\times \bV^*)\circ F(\bU\times \bV
\times \bW)=$$
$$F(\bW^*\times \bU^*\times 1_\sigma)\circ F(\bU\times
\bW\times 1_\sigma)\circ F(1_\rho\times \bW^*\times \bV^*)\circ F(1_\rho\times
\bV\times \bW),$$
and we conclude that
$$\varepsilon(\rho\sigma,\tau)=\varepsilon(\rho,\tau)\times 1_\sigma\circ 
1_\rho \times \varepsilon(\sigma,\tau).$$
Now,
$$\varepsilon^{-1}(\rho,\sigma) \doteq \varepsilon(\sigma,\rho)^{-1}=
F(\bU^*\times \bV^*)\circ F(\bV\times \bU)$$
and as we have here just interchanged the roles of $\cU$ and $\cV$, we
deduce that
$$\varepsilon(\rho,\sigma\tau)=1_\sigma\times \varepsilon(\rho,\tau)\circ 
\varepsilon(\rho,\sigma)\times 1_\tau$$
as stated. \qed

   There still remains the question of whether $\varepsilon$ is in fact a 
{\it symmetry}, i.e. whether $\varepsilon=\varepsilon^{-1}$. But in the above 
proof, we have just seen that  we pass from $\varepsilon$ to 
$\varepsilon^{-1}$ by interchanging the roles
of $\cU$ and $\cV$. Hence one way of tackling the problem is to ask when 
two bi-asymptopias give rise to the same braiding and here we may follow 
our discussion in the case of asymptopias. %\smallskip

Two bi-asymptopias will obviously give rise to the same braiding if 
one is included in the other, or if their
intersection is still a bi-asymptopia; thus if we order the set of
bi-asymptopias under inclusion, we can divide that set into path-components,   
and the braiding associated to a bi-asymptopia  will depend only on
its path-component.%\smallskip 

A difference from the case of asymptopias is worth noting here: 
whilst $\rho\mapsto\bcU_{\r_\cU}$ defines a maximal 
asymptopia, where in particular equation (\ref{2}) is automatically satisfied,
the notion of asymptotic Abelianness entering in the definition of 
bi-asymptopias requires a further mutual (asymptotic) commutativity, 
equation (\ref{3}). Hence 
there is no a priori unique  maximal bi-asymptopia containing a given one, 
just as in general there is no unique maximal Abelian 
subalgebra containing a given Abelian 
subalgebra. Thus in this case we cannot say a priori that 
{\it different} braidings correspond to different path components of 
asymptopias. But by the previous discussion we have\smallskip

\noindent
{\bf Theorem 9} {\sl The bi--asymptopia $\{\cU,\cV\}$ gives rise to a 
symmetry if it lies in the same path--component as 
$\{\cV,\cU\}$.}\smallskip

We close with a few comments. First, when
does a braiding arise from a bi-asymptopia?
Note that if $\varepsilon$ is a
braiding for $\cT$, then this braiding extends in an obvious way   
to a braiding for $\bcT$. Writing $\br$ to denote an
object of $\bcT$, \ie a sequence of elements
of $\Delta$, $\bU\in(\rho,\br)$ and $\bV\in(\sigma,\bs)$, then
$$\bV\times \bU\circ
\varepsilon(\rho,\sigma)=\varepsilon(\br,\bs)
\circ \bU\times \bV.$$
So if we give ourselves a braiding $\varepsilon$ for $\cT$ then
$$\varepsilon(\rho,\sigma)=F(\bV^*\times  \bU^*\circ  \bU\times  \bV)$$
if and only if $F(\varepsilon(\br,\bs))=1_\iota$. Next,
$$\varepsilon(\br',\bs')\circ ( \bU'R \bU^*)\times ( \bV'S \bV^*)=    
( \bV'S \bV^*)\times ( \bU'R \bU^*)\circ \varepsilon(\br,\bs).$$ 
Finally, $F(\br)=F(\br')=F(\bs)=\iota$ and
$F(\varepsilon(\br,\bs))=F(\varepsilon(\br',\bs))=
1_\iota$ implies $F(\br\br')=\iota$ and
$F(\varepsilon(\br\br',\bs))=1_\iota$. 
So we conclude\smallskip 

\noindent
{\bf Theorem 10} {\sl Let $\varepsilon$ be a braiding for $\cT$ and 
suppose given two mappings $\check{\cU},\check{\cV}$ from the
morphisms $\rho$ into stable subsets of $\cU(\bcB)_\rho$
such that each $\check{\bcU}_\rho$ and each $\check{\bcV}_\rho$
are non--empty. If, given any pair $\bU\in(\rho,\br)$ from $\check{\bcU}_\rho$
and $\bV\in(\sigma,\bs)$ from $\check{\bcV}_\sigma$,
$F(\br)=F(\bs)=\iota$
$F(\varepsilon(\br,\bs))=1_\iota$, then $\check{\cU},\check{\cV}$ can be
extended to a bi-asymptopia $\{\cU ,\cV \}$ giving 
$\varepsilon$.
Furthermore, we can even take ${\cU}$ and ${\cV}$ to be closed 
under
composition on the right by unitary intertwiners and stable under tensor
products.}\smallskip

   As a final comment, we note that our notion of asymptotic Abelianness 
of the $\times$--product implies the corresponding notion for the operator
product. In fact, $F(\bU'R\bU^*)=F(\bU'R\bU^*\times 1_{\bs'})$ and
$$F(\bV'S\bV^*)=F(\bV'S\bV^*\times 1_{\br})=F(\bV'S\bV^*\times 
\bU1_\rho  \bU^*).$$  
Hence
$$
 F(\bU'R\bU^*)\circ F(\bV'S\bV^*) 
 = F(\bU'R\bU^* \! \times  \! 1_{\bs'})\circ F(\bU1_\rho 
\bU^*  \! \times  \! \bV'S\bV^*)
$$
when we have a bi-asymptopia. Thus
$$F(\bU'R\bU^*)\circ F(\bV'S\bV^*)=F(\bU'R\bU^*\times \bV'S\bV^*)$$ and
$$F(\bU'R\bU^*)\circ F(\bV'S\bV^*)=F(\bV'S\bV^*)\circ F(\bU'R\bU^*)$$
as stated.

\section{Algebraic Quantum Field Theory}

  We briefly outline the relations of the preceding general
analysis to algebraic quantum field theory, 
giving in particular some examples of asymptopias.

To begin with, 
we consider the semigroup of localized morphisms $\rho$ of an 
observable net $\mathfrak A$ on Minkowski space which are transportable as 
representations on the vacuum Hilbert space of the theory. 
The $C^*$--algebra $\cA$ of Section 2 can then be 
thought of as the $C^*$--inductive limit of 
the net $\fA(\cO)$ of local algebras associated with double cones $\cO$,
$\cA \doteq \overline{\bigcup_\cO \, \fA(\cO)}$. The intertwiners in the 
sense of representations belong to the dual net 
$\mathfrak A^d$, 
$$\fA^d(\cO) \doteq {\textstyle \bigcap}_{\, \cO_1 \subset \cO'} \, 
\fA(\cO_1)',$$
where as usual $\cO'$ denotes the spacelike complement of the double 
cone $\cO$. The role of $\cB$ is played by the 
$C^*$--algebra generated by the net $\mathfrak A^d$, 
$\cB \doteq \overline{\bigcup_\cO \, \fA^d(\cO)}$.
We know that if $U_m\in(\rho,\rho_m)$ is unitary and $\rho_m$ 
is localized in $\cO_m$ then 
$$\rho(A)=U_m^*AU_m,\quad A\in\mathfrak A(\cO),\quad \cO_m\subset \cO'.$$
Hence our endomorphisms are asymptotically inner in $\cB$. If we 
assume Haag duality we have $\cA=\cB$, so 
we do not need to extend our endomorphisms 
and $\rho\mapsto\cU(\bcB)_\rho$ is the unique maximal asymptopia.

  If we just assume essential duality, then in space--time 
dimension $d>2$, we know that if 
$R_i\in(\rho_i,\sigma_i)$, $i=1,2$ then $R_1R_2=R_2R_1$ if $\rho_1$ 
and $\sigma_1$ 
are localized spacelike to $R_2$. Thus we can define an asymptopia $\cU$
by taking $\bcU_\rho$ to consist of all unitaries of 
$\cU(\bcB)$ with 
representing sequences $\bU_m\in(\rho,\rho_m)$, the $\rho_m$ being 
localized in double cones $\cO_m$ tending spacelike to infinity. 
In this case, $\cA_\D=\cB$ and our Theorem 2 is just a variant of 
a known result cf.\ \cite[\S 3.4.6]{R}.

The reader's attention is also drawn to the case of charges localizable 
in spacelike cones \cite{BF} where the intertwiners likewise do not lie in 
the observable 
algebra and where the treatment in \cite{DR} has aspects in common with the 
present paper, cf.\ Lemma 5.5 of \cite{DR}.

   In space--time dimension $d=2$, 
the spacelike complement of a double cone has two path--components, 
a spacelike left and a spacelike right. 
As far as the commutation properties of intertwiners 
$R_i\in(\rho_i,\sigma_i)$ go, we merely know then 
that $R_1R_2=R_2R_1$ 
if $\rho_1$ and $\sigma_1$ are both localized left spacelike 
to $R_2$ or both localized  right spacelike 
to $R_2$. Thus we can define two asymptopias $\cU^\ell$ and $\cU^r$ 
as above by 
letting $\cO_m$ tend spacelike to left infinity or right infinity, 
respectively. The restricted commutation properties of intertwiners 
show that these two asymptopias lead to different tensor structures, 
in general. The vacuum representation of the observable net then 
induces a solitonic representation of the corresponding field net.

In replacing $\mathfrak A$ by the algebra $\cA$, we are, on the one
hand, simplifying the mathematical setting 
by suppressing the net structure and avoiding all reference to spacetime. 
But we also have in mind applications where the endomorphisms are no 
longer strictly localized but only asymptotically inner. 

  For an example going beyond the standard setting of strictly
local or cone--localized charges, we turn to the model expounded in
\cite{BDMRS1} and based on the free massless scalar field. $\cA$  is
here the $C^*$--algebra generated by Weyl
operators $W(f)$, $f\in\cL$,
$$\cL\doteq \omega^{-\frac{1}{2}} \cD(\mathbb R^3)+i\omega^\frac{1}{2}
\cD(\mathbb R^3).$$
Here $\cD(\mathbb R^3)$ denotes the space of smooth real-valued
functions with compact support and $\omega$ the energy operator. 
$\cL$ is equipped with the scalar product
$$(f,f') \doteq \int \! d^3\bx\overline{f(\bx)}f'(\bx), $$
determining the symplectic form 
$$\sigma(f,f')=-\Im (f,f')$$
and the usual vacuum state:
$$\omega(W(f))=e^{-\frac{1}{4}(f,f)}.$$
The $C^*$--algebra $\cB$ can be taken to be the algebra of 
all bounded operators on the vacuum Hilbert space. For $\Delta$, we
take the group $\Gamma$ of automorphisms of $\cA$ generated by the space
$$\cL_\Gamma \doteq \omega^{-\frac{1}{2}}\cD(\mathbb R^3)+
i\omega^{-\frac{3}{2}}\cD(\mathbb R^3).$$
For convenience, we use the same symbol $\gamma$ to denote the element
of $\cL_\Gamma$, parametrized by smooth functions $g$ and $h$,
$$\gamma=i\omega^{-\frac{3}{2}}g+\omega^{-\frac{1}{2}}h,\quad
g,h\in\cD(\mathbb R^3),$$  
and the automorphism it generates so that
$$\gamma(W(f))=e^{i\sigma(\gamma,f)}W(f),$$
where the symplectic form $\sigma$ is defined on $\cL_\Gamma$ so as
to extend that on the subspace $\cL$ by
$$\sigma(\gamma,\gamma')=\int \! d^3\bp\,\omega^{-2}\big(\tilde g(-\bp)
\tilde h'(\bp)-\tilde g'(-\bp)\tilde h(\bp)\big).$$

The sectors are characterized by the charge
$$\int \! d^3\bx \, g(\bx).$$ They are translation invariant
and if $\gamma_a$ denotes the translate of $\gamma$ by $a$, we have
unitary intertwiners $U_a\in(\gamma,\gamma_a)$ unique up to a phase.
We define $\bcU_\gamma$ to be the set of equivalence 
classes of sequences of unitaries
$U_a\in(\gamma,\gamma_a)$ for which $a$ tends spacelike to infinity 
and $a_0 / |\ba| \to 0$. Then
$$U_a^*W(f)U_a=e^{i\sigma(\gamma-\gamma_a,f)}W(f)\to\gamma(W(f)),$$
as follows from the asymptotic behaviour of the symplectic form,
Theorem 3 of \cite{BDMRS1}. Obviously,
$\bcU_\gamma\bcU_\delta=\bcU_{\delta\gamma}$. Hence to show that
the assignment 
$\gamma\mapsto \bcU_\gamma$ defines an asymptopia, it suffices to check that
the intertwiners are asymptotically Abelian. Now if
$(\gamma,\delta)\neq 0$, it consists of multiples of $W(\delta-\gamma)$.
Hence the intertwiners are asymptotically Abelian if
$$\lim_{a,b} \, [W(\delta_b-\gamma_a),W(\delta'-\gamma')]=0,$$
whenever $\gamma$ and $\delta$ are equivalent and $\gamma'$ and
$\delta'$ are equivalent. The norm of this commutator of Weyl
operators is
$$|e^{i\sigma(\delta_b-\gamma_a,\delta'-\gamma')}-1|$$
and asymptotic Abelianness follows from Theorem 3 of \cite{BDMRS1}.

In order to define a bi-asymptopia, we may take $\bcU_\gamma$ to consist of 
representing sequences
$U_a\in(\gamma,\gamma_a)$, where $a$ tends to spacelike infinity
inside some (open) spacelike cone $\cS$, and define $\bcV_\gamma$ 
similarly
using the spacelike cone $-\cS$. In view of our previous computations,
we need only verify the condition of
asymptotic Abelianness. Since every non-zero element of
$(\gamma,\gamma')$
is a multiple of $W(\gamma'-\gamma)$, it suffices to check that
$$W(\gamma'_{a'}-\gamma_a)\times W(\delta'_{b'}-\delta_b)-
W(\delta'_{b'}-\delta_b)\times W(\gamma'_{a'}-\gamma_a) $$
tends to zero as $a$, $a'$, $-b$ and $-b'$ go spacelike to infinity in 
$\cS$.
But the norm of this expression is
$$|e^{i\sigma(\gamma'_{a'},\delta'_{b'})}e^{i\sigma(\delta_b,\gamma_a)}-1|$$
and tends to zero as required by Theorem 3 of \cite{BDMRS1}.

  It is easy to see by direct computation that the braiding determined
by this bi-asymptopia is given by 
$$\varepsilon(\gamma,\delta)=e^{-i\sigma(\gamma,\delta)},$$ 
cf.\ Sect.\ 5 of \cite{BDMRS1}, and is therefore a symmetry, but
it is instructive to derive this from Theorem 9. Obviously, if we
replace the spacelike cone $\cS$ in the definition of the 
bi-asymptopia
by a smaller spacelike cone $\cS_1$ we remain in the same 
path--component.
The same is therefore true if $\cS\cap\cS_1\neq\emptyset$. By a
sequence of such moves, we may interchange $\cS$ and $-\cS$ so that by
Theorem 9 our braiding is a symmetry.%\smallskip

\bigskip

\noindent {\bf \Large Acknowledgements} \\[2 mm] 
SD would like to thank the II.\ Institut f\"ur Theoretische Physik,
Universit\"at Hamburg, and SD and JER would like to thank the 
Institut f\"ur Theoretische Physik, Universit\"at G\"ottingen, 
for their kind hospitality in the final stage of this 
collaboration. Both gratefully acknowledge the financial support
of the Alexander von Humboldt Foundation, that made this collaboration
possible.

\end{document}